\documentstyle[12pt]{article}

\newcommand{\be}{\begin{equation}}
\newcommand{\ee}{\end{equation}}
\newcommand{\bear}{\begin{eqnarray}}
\newcommand{\ear}{\end{eqnarray}}
\date{}

\newcommand{\GeV}{\mbox{$\;$GeV}}
\newcommand{\bm}[1]{\mbox{\bf #1}}  
\renewcommand{\vec}{\bm}
\newcommand{\grgl}{\:\hbox to -0.2pt{\lower2.5pt\hbox{$\sim$}\hss}
           {\raise3pt\hbox{$>$}}\:}
\newcommand{\klgl}{\:\hbox to -0.2pt{\lower2.5pt\hbox{$\sim$}\hss}
           {\raise3pt\hbox{$<$}}\:}
\begin{document}
\begin{titlepage}
\begin{flushright}
HD--THEP--94--42
\end{flushright}
\vspace{1.8cm}
\begin{center}
{\bf\LARGE NONPERTURBATIVE QCD EFFECTS}\\
\bigskip
{\bf\LARGE  IN HIGH ENERGY COLLISIONS}\\
\vspace{1cm}
by\\
\vspace{1cm}
O. Nachtmann\\
\bigskip
Institut  f\"ur Theoretische Physik\\
Universit\"at Heidelberg\\
Philosophenweg 16, D-69120 Heidelberg, FRG\\
\vspace{3cm}
{\bf Abstract:}\\
\parbox[t]{\textwidth}{High energy hadron-hadron collisions are discussed.
It is argued that soft collisions should involve in an essential way
nonperturbative
QCD. A way is outlined how to calculate properties of high energy elastic
hadron-hadron scattering using field theoretic methods. The functional
integrals
occuring there are  evaluated using the ``stochastic vacuum model''. A
satisfactory comparison between theory and experiment is achieved. Then the
question
of possible nonperturbative QCD effects in high energy hard hadron-hadron
collisions
is raised. It is shown that  some spin effects in the Drell-Yan process may
give a
hint that such effects exist indeed in nature. }
\end{center}
\vfill
\noindent Talk presented at the 18th Johns Hopkins Workshop, Florence, 1994
\end{titlepage}
\newpage
\section{Introduction}
\setcounter{equation}{0}
Today, most of us will agree that all the phenomena of hadron physics
should be described in the framework of QCD.  We know the simple
and elegant Lagrangian density of QCD, ${\cal L}_{\rm QCD}$, and
in principle
everything is derivable from it. We have
\be\label{1.1}
{\cal L}_{\rm QCD}(x)=-\frac{1}{4} G^a_{\lambda\rho}(x) G^{a\lambda
\rho}(x)
+\sum_q \bar q(x)(i\gamma^\lambda D_\lambda-m_q)q(x),\ee
where $G^a_{\lambda\rho}(x)(a=1,\ldots,8)$ are the components of the gluon
field strength tensor, $q(x)$ are the quark fields, $m_q$ the quark masses and
$D_\lambda$ the covariant derivative. (All our notation follows \cite{1}).

But the degrees of freedom in the Lagrangian density are quarks and gluons,
not the hadrons we observe in nature. To make quantitative predictions for the
real
world starting from ${\cal L}_{\rm QCD}$ is not easy. In two areas this has
been
successful:
\begin{enumerate}
\item for short distance  phenomena, where due to \underbar{asymptotic freedom}
perturbation theory can be applied;
\item for hadron spectroscopy and other long distance phenomena, where
numerical,
 nonperturbative methods can be applied, e.g. Monte Carlo simulations on a
lattice or
the discretized light cone quantization procedure.
\end{enumerate}

There is a third class of phenomena which are neither pure short distance nor
pure long distance: high energy hadron-hadron collisions which have contributed
so
much to our understanding of hadronic phenomena. These reactions are
traditionally
classified into ``hard'' and ``soft'' ones:
\begin{enumerate}
\item[3.] High energy hadron-hadron collisions:
\begin{itemize}
\item[(3.a)] hard reactions,
\item[(3.b)] soft reactions.
\end{itemize}
\end{enumerate}

Typical hard reactions are the Drell-Yan-type processes, e.g.
\bear\label{1.2}
\pi^-+N\to&&\gamma^*+X,\nonumber\\
&&\hookrightarrow\ell^+\ell^-
\ear
where $\ell=e,\mu$. All energies and momentum transfers are assumed to be
large.
However, the masses of the $\pi^-$ and $N$ in the initial state stay fixed and
thus we are not dealing with a pure short distance phenomenon.

In the reaction (\ref{1.2}) we claim to see directly the fundamental quanta of
the
theory, the partons, i.e. the quarks and gluons, in action (cf. Fig. 1). In the
usual theoretical  framework for hard reactions, the QCD improved parton model
(cf.
e.g. \cite{2}), one describes the reaction of the partons, in the Drell-Yan
case the
$q\bar q$ annihilation into a virtual  photon, by perturbation theory. This
should be
reliable, since the parton process involves only high energies and high
momentum
transfers. All the long distance physics due to the bound state nature of the
hadrons is then lumped into parton distribution functions of the participating
hadrons. This is called the \underbar{factorization hypothesis}, which after
early
investigations of soft initial and final state interactions \cite{3} was
formulated
and studied in low orders of QCD perturbation theory in \cite{4}. Subsequently,
great theoretical effort has gone into proving factorization in the framework
of QCD
perturbation theory \cite{5}-\cite{7}. The result seems to be that
factorization is most
probably correct there.  However, it is legitimate to ask if factorization is
respected also by nonperturbative effects.  To my knowledge this question was
first
asked in \cite{8}-\cite{10}. In Sect. 4 of this seminar I will come back to
this question
and will argue that there may be evidence for a breakdown of factorization in
the
Drell-Yan reaction.

The other class of reactions I will discuss in the  following  are soft high
energy
collisions. A typical reaction we will be interested in is proton-proton
elastic
scattering:
\be\label{1.3}
p+p\to p+p\ee
at c.m. energies $E_{\rm cm}=\sqrt{s}\grgl 5{\GeV}$ say and small momentum
transfers $\sqrt{|t|}=|q|\klgl1{\GeV}$. Here we have two scales, one staying
finite, one going to infinity:
\bear\label{1.4}
E_{\rm cm}&&\to \infty,\nonumber\\
|q|&&\klgl 1{\GeV}.\ear
Thus, none of the above mentioned calculational  methods is directly
applicable.
Indeed, most theoretical papers dealing with reactions in this class  develop
and
apply \underbar{models} which are partly older than QCD, partly QCD
``motivated''.
Let me list \underbar{some models for hadron-hadron elastic scattering} at high
energies:

geometric \cite{11},

eikonal \cite{12},

additive quark model \cite{13},

Regge poles \cite{14},

topological expansions and strings \cite{15},

valons \cite{16},

leading log summations \cite{17},

two-gluon exchange \cite{18},

etc.

It would be a forbidding  task to collect all references  in this field. The
references given above should thus only be considered as representative ones.
In
addition I would like to mention the inspiring general field theoretic
considerations for high energy scattering and particle production by Heisenberg
\cite{19} and the impressive work by Cheng and Wu on high energy behaviour in
field
theories in the framework of perturbative calculations \cite{20}.

I will now argue that the theoretical description of measurable  quantities of
soft
high energy reactions like the total cross sections should involve in an
essential
way \underbar{nonperturbative} QCD.
To see this,  consider massless pure gluon theory where all ``hadrons'' are
massive glueballs. Then we know from the renormalization group analysis that
the
glueball masses must behave as
\be\label{1.5}
m_{\rm glueball}\propto M e^{-c/g^2(M)}\ee
for $M\to\infty$, i.e. for $g(M)\to 0$, due to asymptotic freedom. Here $M$  is
the
renormalization scale, $g(M)$ is the QCD coupling strength at this scale and
$c$ is
a constant. Masses in massless Yang-Mills theory are a purely nonperturbative
phenomenon, due to ``dimensional transmutation''. Scattering of
glueball-hadrons in
massless pure gluon theory should look very similar to scattering of hadrons in
the
real world,  with finite total cross sections, amplitudes with analytic $t$
dependence etc. At least, this would be my expectation. If the total cross
section
$\sigma_{\rm tot}$ has a finite limit as $s\to\infty$ we must have from the
same
renormalization group arguments:
\be\label{1.6}
\lim_{s\to\infty}\sigma_{\rm tot}(s)\propto M^{-2} e^{2c/g^2(M)}\ee
for $g(M)\to 0$.
In this case, the total cross sections in pure gluon theory are also
nonperturbative
objects! It is easy to see that this conclusion is not changed if $\sigma_{\rm
tot}(s)$ has a logarithmic behaviour with $s$ for $s\to\infty$, e.g.
\be\label{1.7}
\sigma_{\rm tot} (s)\to {\rm const} \times (\log s)^2.\ee
I would then expect that also in full QCD  total cross sections are
nonperturbative
objects, at least as far as hadrons made out of light quarks  are concerned.

Some time ago  P.V. Landshoff   and myself started to think about a possible
connection between the nontrivial vacuum structure of QCD - a typical
nonperturbative phenomenon - and soft high energy reactions \cite{21}.  In the
following  I will first review  some common folklore on the QCD vacuum and then
sketch possible consequences of these ideas for high energy collisions.

\section{The QCD Vacuum}
\setcounter{equation}{0}

According to current theoretical  prejudice  the vacuum state in QCD has a very
complicated  structure \cite{22}-\cite{32}. It was first noted by Savvidy
\cite{22}
 that by
introducing a constant chromomagnetic  field
\be\label{2.1}
\vec B^a=\vec n\eta^a B,\ (a=1, \ldots, 8),\ee
into the \underbar{perturbative} vacuum one can lower the vacuum-energy density
$\varepsilon(B)$. Here $\vec n$ and $\eta^a$ are constant unit vectors in
ordinary
and colour  space.  The result of his one-loop calculation was
\be\label{2.2}
\varepsilon(B)=\frac{1}{2}B^2+\frac{bg^2}{32\pi^2}
B^2\left[\ln\frac{B}{M^2}-\frac{1}{2}\right]\ee
where $g$ is the strong coupling constant, $M$ is again the renormalization
scale,
and $b$ is given by the lowest order term in the Callan-Symanzik
$\beta$-function:
\be\label{2.3}
\beta(g)=-\frac{b}{16\pi^2}g^3+\ldots\ee
For 3 colours and $f$ flavours:
\be\label{2.4}
b=11-\frac{2}{3} f\ee
thus, as long as we have asymptotic  freedom, i.e. for $f\leq 16$, the energy
density $\varepsilon(B)$ looks as indicated schematically in Fig. 2 and has its
minimum for $B=B_{\rm vac}\not=0$. Therefore, we should expect the QCD-vacuum
to
develop spontaneously a chromomagnetic field, the situation being similar to
that in
a ferromagnet below the Curie temperature where we have spontaneous
magnetization.

Of course, the vacuum state in QCD has to be relativistically invariant and
cannot
have a preferred direction in ordinary space and colour space. What has been
considered \cite{28}  are states composed of domains with random orientation of
the
gluon-field strength (Fig. 3). This is analogous  to Weiss domains in a
ferromagnet.  The vacuum state should then be a suitable linear superposition
of
states with various domains and orientation of the fields inside the domains.
This
implies that the orientation of the fields in the domains as well as the
boundaries
of the domains will fluctuate.

A very detailed picture for the QCD vacuum along these lines has been developed
in
ref. \cite{28}. I cannot refrain from comparing this modern picture of the QCD
vacuum (Fig. 4a) with the ``modern picture'' of the ether developed by
\underbar{Maxwell} more than 100 years ago (Fig. 4b). The analogy is quite
striking
and suggests to me that with time passing on we may also be able to find
simpler
views on the QCD vacuum as Einstein did with the ether. In the following we
will
adopt the picture of the QCD vacuum as developed in refs.
 \cite{22}-\cite{29},\cite{31} and
outlined above as a \underbar{working hypothesis}.

Let me now come to the values for the field strengths $\vec E^a$ and $\vec B^a$
in
the vacuum. These must also be determined by $\Lambda$, the QCD scale
parameter, the
only dimensional parameter in QCD if we disregard  the quark masses. Therefore,
we
must have on dimensional grounds for the renormalization group invariant
quantity
$(gB)^2$
\be\label{2.5}
(gB)^2\sim\Lambda^4.\ee
But we have much more detailed information on the values of these field
strengths
due to the work of Shifman, Vainshtein, and Zakharov (SVZ), who introduced the
\underbar{gluon condensate} and first estimated its value using sum rules for
charmonium  states \cite{25}:
\bear\label{2.6}
<0|\frac{g^2}{4\pi^2} G^a_{\mu\nu}(x)G^{\mu\nu a}(x)|0>&\equiv
&<0|\frac{g^2}{2\pi^2}\left(\vec B^a(x)\vec B^a(x)-\vec E^a(x)\vec
E^a(x)\right)|0>\nonumber\\
&\equiv &G_2= (2.4\pm 1.1)\cdot 10^{-2} {\rm GeV}^4\nonumber\\
&=&(335-430 {\rm MeV})^4.\ear
Here we quote numerical values as given in the review \cite{34}.
A simple analysis shows that this implies
\be\label{2.7}
<0|g^2\vec B^a(x)\vec B^a(x)|0>=-<0|g^2\vec E^a(x)\vec E^a(x)|0>=\pi^2
G_2\simeq(700
{\rm MeV})^4.\ee
To prove eq. (\ref{2.7}) w note that Lorentz- and Parity-invariance require the
vacuum expectation value  of the uncontracted product of two gluon field
strengths
to be of the form
\be\label{2.8}
<0|\frac{g^2}{4\pi^2} G^a_{\mu\nu}(x) G^b_{\rho\sigma}(x)|0>=(g_{\mu\rho}
g_{\nu\sigma}-g_{\mu\sigma} g_{\nu\rho})\delta^{ab}\frac{G_2}{96}\ee
where $G_2$ is the same constant as in (\ref{2.6}). Taking appropriate
contractions leads to (\ref{2.6}) and (\ref{2.7}).

We find that $<0|\vec B^a(x)\ \vec B^a(x)|0>$ is positive, $<0|\vec E^a(x)\
\vec E^a(x)|0>$ negative! This can happen because we are really considering
products of field operators, normal-ordered with respect to the perturbative
vacuum. The interpretation of eq. (\ref{2.7}) is, therefore, that the B-field
fluctuates with bigger amplitude, the E-field with smaller amplitude
than in the perturbative vacuum state.

What about the size $a$ of the colour domains and the fluctuation times $\tau$
of the colour fields? On dimensional grounds we must have
\be\label{2.9} a\sim \tau\sim\Lambda^{-1}.\ee
A detailed model for the QCD vacuum incorporating the gluon condensate idea
and a fall-off of the correlation of two field strengths with distance
was proposed in \cite{35}: the ``stochastic vacuum model''. There the
basic object is the vacuum expectation value of the ``gluon field
strength correlator'':
\be\label{2.1a} <0|G^a_{\mu\nu}(x)\ {\rm string\ operator}\
G^b_{\rho\sigma}(y)|0>.\ee
This is the vacuum expectation value of the product of two field strengths
at different points $x,y$ made gauge-invariant by the insertion of a
suitable Schwinger string operator \cite{37}, \cite{34}. Now an
ansatz for the function (\ref{2.1a}) is made. The fall-off of this
function as $|x-y|$ gets large can be used to give a precise meaning
to $a$ (\ref{2.9}) as a correlation length. Various measurable quantities
can be calculated within this model, and the parameters of the
above-mentioned ansatz can be determined by comparison with experiment. In
this way one finds from \underbar{low energy} phenomenology \cite{34}
\be\label{2.10} a\simeq 0.35\ {\rm fm}.\ee
This is smaller, albeit not much smaller, than a typical radius $R$ of a
light hadron (cf. e. g. \cite{36}):
\be\label{2.11}R\sim 0.7-1\ {\rm fm}.\ee
Still
\be\label{2.12}
a^2/R^2\approx 0.2-0.3\ee
is a reasonably small number and this will be important for us in the
following.

The stochastic vacuum model is quite interesting as it leads for
instance to an easy and intuitive understanding of confinement \cite{35}.
For us here it will be a tool to evaluate functional integrals in the
nonperturbative domain.

\section{Soft Hadronic Reactions}
\setcounter{equation}{0}
Consider now a quark sailing through the vacuum. The quark will interact with
the gluons in the vacuum, it will have a gluon cloud around it, where the
strength of the gluon cloud is governed by $g^2<G_{\mu\nu}G^{\mu\nu}>$
and its extension in transverse directions by $a$. Suppose a second quark
comes in opposite direction. The interaction of the two quarks happens
in essence in the following way: one quark changes the gluon distribution
in the vacuum, this change is felt by the other quark. Thus we estimate
naively that in this picture two quarks must come closer than a distance
of order $a$ in transverse directions for scattering to occur (Fig. 5).

This quark-quark encounter could, for instance, take place in a
$pp$-elastic scattering at high energies (Fig. 6). since $a^2\ll R^2$
(\ref{2.12}), the chance to find two quarks of \underbar{one} proton
simultaneously at a transverse distance $\klgl a$ of a quark in the
other proton is small. We arrive at the conclusion that single
quark scattering should dominate! This would be a simple explanation
of the success of the additive quark rule \cite{13} found nearly
30 years ago. Note that it is very difficult to understand this rule
in perturbative QCD where gluon exchange leads in essence to a
long-range Coulomb-type force where formally $a\to\infty$.

In \cite{21} P. V. Landshoff and myself investigated an abelian
gluon model which realized the above simple physical picture.
In \cite{38} I developed these ideas further, starting from
 ${\cal L}_{QCD}$ and trying
to use only ``honest'' field-theoretic methods.  I could show that
the quark-quark scattering amplitude can be obtained by first
calculating the amplitudes for each quark $q_1,q_2$ scattering in
an external gluon potential (Fig. 7). Let the amplitudes of this
scattering be $M_1(G),M_2(G)$ for the quarks $q_1,q_2$, respectively.
In a second step one has to average the product of these two scattering
amplitudes over all gluon potentials with an appropriate functional
integral measure:
\be\label{3.1}<q_1q_2|T|q_1q_2>\ \propto\ <M_1(G)M_2(G)>_{G-average}.\ee
Furthermore I found that in the high-energy limit $M_{1,2}(G)$
are governed by the nonabelian phase factors
\be\label{3.2}
V_{1,2}(G)\sim P\left\lbrace\exp-ig\int_{1,2}dz_\mu G^\mu(z)\right\rbrace,
\ee
where
\be\label{3.3}
G^\mu(z)=G^{a\mu}(z)\frac{\lambda_a}{2}\ee
is the gluon potential matrix, and $P$ means path ordering.
The line integrals run along the paths travelled by the quarks
$q_{1,2}$ in Minkowski space. Inserting this in (\ref{3.1}) leads to
\be\label{3.4}
<q_1q_2|T|q_1q_2>\ \sim\ <V_1(G)V_2(G)>_{G-average}.\ee

How can we discuss such an expression further? Equation (\ref{3.4})
means that we have to take a whole sum of gluon interactions of two
quark lines and average them (Fig. 8). Surely we do not want
to go back to the perturbative expansion in powers of $g$ from
there, since we remember our argument (\ref{1.6}). Instead we can
make a nonperturbative ansatz for the gluon propagator and in an
abelian model reproduce the results of \cite{21} in an easy way. To
generalize this for full QCD with its non-abelian character we will
use the methods of the stochastic vacuum model. The
original version of this model was formulated by Dosch and Simonov
for functional integrals in Euclidian space \cite{35}.
However, to evaluate the functional integral on the r.h.s. of (\ref{3.4})
we have to work in Minkowski space --- at least we found no way to
make a sort of Wick rotation in this case. To deal with such a
situation, Dosch and Kr\"amer \cite{39} extended the stochastic
vacuum model. They made an analytic continuation --- i.e. a Wick
rotation --- for the correlators (\ref{2.1a}), originally defined in
Euclidian space, from there to Minkowski space. They also introduced
a suitable factorization scheme for
higher point  correlators of gluon field strengths. These ans\"atze
give us now a method to evaluate functional integrals in Minkowski space
in the nonperturbative domain. The contributions to such functional
integrals can be ordered according to the number of 2-point correlators
appearing.

For $qq$-elastic scattering with no net colour exchange the simplest
contribution involves two 2-point gluon correlators and
has a structure reminiscent of 2-gluon exchange (Fig. 9).
Thus we expect for the amplitude
\bear\label{3.5}
<qq|T|qq>|_{elastic}
&\propto&<0|G\ G\ G\ G|0>\nonumber\\
&\propto&<0|G\ G|0><0|G\ G|0>\nonumber\\
&\propto&G_2^2,\ear
where $G_2$ is the gluon condensate and we have used the assumption of
factorization for the 4-point correlator. Our estimate for the
total $qq$ cross section is then through the optical theorem
\be\label{3.6}
\sigma_{tot}(qq)\propto const.\ G_2^2.a^{10}.\ee
Here we have given the correct dimension to $\sigma_{tot}$ by
multiplying with the appropriate power of the vacuum correlation
length $a$, the only dimensionful quantity available in this problem
apart from $G_2$ if we set the quark masses to zero. Through the
additive quark rule we estimate also:
\be\label{3.7}
\sigma_{tot}(pp)=const. \ G^2_2.a^{10}.\ee

The explicit calculations in the framework of the Minkowskian version
of the stochastic vacuum
model lead to the following results \cite{39}-\cite{41}. For an abelian gluon
theory the above estimates (\ref{3.6}), (\ref{3.7}) are correct, the additive
quark rule for total cross sections at high energies holds. For the
non-abelian theory the additive quark rule is \underbar{not} obtained
in this framework. It turns out that the amplitude for single quarks
scattering on each other is not a sensible object. Sensible objects are the
amplitudes for scattering of mesons, considered as $q\bar q$ wave packets,
on each other. Similarly, the scattering of mesons or baryons on baryons
can be treated as scattering of $q\bar q$ or $qqq$ on $qqq$ wave
packets. Definite rules for writing down and evaluating these scattering
amplitudes can be given. Now, in addition to the vacuum parameters $G_2$ and
$a$
also the radii of the hadrons --- i.e. of the wave packets representing
them --- enter in the results. A fit to numerical results gives for the
total cross section and the slope parameter at $t=0$ of elastic proton-proton
scattering
\be\label{3.8}
\sigma_{tot}(pp)=0.00881\left(\frac{R_p}{a}\right)^{3.277}\cdot
(3\pi^2G_2)^2.a^{10},\ee
\be\label{3.9}
b_{pp}:=\frac{d}{dt}\ln\frac{d\sigma_{el}}{dt}(pp)|_{t=0}=1.558 a^2+0.454
R^2_p.
\ee
Here $R_p$ is the proton radius and the formulae (\ref{3.8}), (\ref{3.9})
are valid for
\be\label{3.10}
1\leq R_p/a\leq 3.\ee

To compare (\ref{3.8}), (\ref{3.9}) with experimental results, we can,
for instance, consider the c.m. energy $\sqrt s=20$ GeV and take as
input the following measured values (cf. \cite{41}):
\bear\label{3.11}
\sigma_{tot}(pp)|_{\rm Pomeron\ part}&=&35\ {\rm mb},\nonumber\\
b_{pp}&=&12.5\ {\GeV}^{-2},\nonumber\\
R_p&\equiv &R_{p,elm}=0.86\ {\rm fm}.\ear
We obtain then from (\ref{3.8}) and (\ref{3.9}):
\bear\label{3.12}
a&=&0.31\ {\rm fm},\nonumber\\
G_2&=&6.61\times 10^{-2}\ {\GeV}^{4}.\ear
The correlation length $a$ comes out in surprising good
agreement with the determination of this quantity from low
energy phenomenology (\ref{2.10}). The gluon condensate value in
(\ref{3.12}) is somewhat larger than the value from sum rule
determinations (\ref{2.6}). There are excuses for that (cf.
\cite{41}).

But perhaps we were lucky in picking out the right c.m. energy $\sqrt s$
and radius for our comparison of theory and experiment. What about
the $s$-dependence of the total cross section $\sigma_{tot}$ and slope
parameter $b$? The vacuum parameters $G_2$ and $a$ should be
independent of the energy $\sqrt s$. On the other hand, it seems quite
plausible to us that the effective strong interaction radii
$R$ of hadrons may depend on $\sqrt s$. Let us consider again $pp$
(or $p\bar p$) elastic scattering. Once we have fixed $G_2$ and $a$
from the data at $\sqrt s=20\GeV$ (\ref{3.8}) and (\ref{3.9}) give us
$\sigma_{tot}(pp)$ and $b_{pp}$ in terms of the single parameter
$R_p$, i.e. we obtain as prediction of the model a curve in the
plane $b_{pp}$ versus $\sigma_{tot}(pp)$. This is shown in Fig. 10. It
is quite remarkable that the data from $\sqrt s=20\GeV$ up to Tevatron
energies, $\sqrt s=1.8$ TeV follow this curve. For more details and
further results we refer to \cite{41}.

Summarizing this section we can say that explicit calculations for high
energy-elastic hadron-hadron scattering near the forward direction have
been performed combining the field-theoretic methods of \cite{38} and the
Minkowski version of the
stochastic vacuum model of \cite{39}. The results are encouraging and
support the idea that the vacuum structure of QCD plays an essential
role in soft high-energy scattering.

\section{``Synchrotron
Radiation'' from the Vacuum and Spin Correlations
in the Drell-Yan Reaction}
\setcounter{equation}{0}
Let us consider for definiteness again a proton-proton collision at high
c.m. energy $\sqrt s\gg m_p$. We look at this collision in the c.m. system
and choose as $z$-axis the collision axis (Fig. 11). According to
Feynman's parton dogma \cite{411} the hadrons look like jets of almost
non-interacting partons, i.e. quarks and gluons. Accepting our
previous views of the QCD vacuum (Sect. 2), these partons travel
in a background chromomagnetic field.

What sort of new effects might we expect to occur in this situation?
Consider for instance a quark-antiquark collision in a chromomagnetic
field. In our picture this is very similar to an electron-positron
collision in a storage ring (Fig. 12). We know that in a storage ring
$e^-$ and $e^+$ are deflected and emit synchrotron radiation. They also
get a transverse polarization due to emission of spin-flip synchrotron
radiation \cite{42}, \cite{43}. Quite similar we can expect the quark
and antiquark to be deflected by the vacuum fields. Since quarks have
electric and color charge, they should then emit both \underbar{photon}
and \underbar{gluon ``synchrotron radiation''}. Of course, as long as
we have quarks within a single, isolated
proton (or antiproton) travelling through the vacuum
no emission of photons can occur, and we should
consider such processes as contributing to the cloud of quasi-real photons
surrounding a fast-moving proton. (This is similar in spirit to the
well-known Weizs\"acker-Williams approximation.) But in a collision
process the parent quark or antiquark will be scattered away and the
photons of the cloud can become real, manifesting themselves as
\underbar{prompt photons in hadron-hadron collisions}.

In ref. 10 we have given an estimate for the rate and the spectrum
of such prompt photons using the classical formulae for synchrotron
radiation \cite{43}. The result we found can be summarized as follows:
In the overall c.m. system of the hadron-hadron collision ``synchrotron''
photons should appear with energies $\omega<300-500$ MeV, i.e. in the
very central region of the rapidity space. The number of photons per
collision and their spectrum are --- apart from logarithms --- independent
of the c.m. energy $\sqrt s$. The dependence of the cross section on the
photon energy and on the emission angle $\theta^*$ with respect to the
beam axis is estimated roughly as follows:
\be\label{4.1}
\frac{1}{\sigma}\frac{d\sigma}{d\omega d\cos\theta^*}\propto\frac{1}
{\omega^{1/3}(\sin\theta^*)^{2/3}}\ee
This should be compared to the inner-bremsstrahlungs spectrum
\be\label{4.2}
\left.\frac{1}{\sigma}\frac{d\sigma}{d\omega d\cos\theta^*}\right|_{\rm
bremsstr.}\propto\frac{1}{\omega\sin^2\theta^*}\ee
The ``synchrotron'' radiation from the quarks should thus be harder
than the brems\-strahlung spectrum. This is welcome, since for $\omega
\to 0$ bremsstrahlung should dominate according to Low's theorem \cite{44}.
It is amusing to note that in several experiments an excess of soft
prompt photons over the bremsstrahlung calculation has been observed
\cite{45}-\cite{49}. The gross features and the order of magnitude of
this signal make it a candidate for our ``synchrotron'' process. A
detailed comparison with our formulae has been made recently at least
for the results from one experiment \cite{49} with encouraging results
\cite{50}.

One might think --- maybe rightly --- that these ideas are a little
crazy. But we have also worked out some consequences of them for
the Drell-Yan reaction (\ref{1.2}), which make us optimistic. In the
lowest order parton process contributing there,  we have a quark-antiquark
annihilation giving a virtual photon $\gamma^*$, which decays then into
a lepton pair (Fig. 1):
\be\label{4.3}
q+\bar q\to \gamma^*\to \ell^+\ell^-.\ee
In the usual theoretical framework $q$ and $\bar q$ are assumed
to be uncorrelated and
unpolarized in spin and colour if the original hadrons are unpolarized.
 From our ideas we would expect a transverse spin correlation due to the
``storage ring'' effect \cite{42}, \cite{43}. We worked this out and found that
this influences the $\ell^+\ell^-$ angular distribution in a profound way. Then
our colleague H. J. Pirner pointed out to us that data which may be
relevant in this connection existed already \cite{51}. And very obligingly
these data seem to support the idea of spin correlations and
thus vacuum effects in high energy collisions. For more details we refer
to \cite{52}. If such spin correlations are confirmed by experiments
at higher energies, we would presumably have to reconsider the
fundamental factorization hypothesis for hard reactions which we sketched
in Sect.~1.

\section{Conclusions}
We have given a sketch of methods and results concerning possible
nonperturbative effects in high energy hadronic collisions. I think
it is rather sure that such effects play a decisive role in soft
hadronic collisions. I find it even more exciting that they may also
influence hard reactions like the Drell-Yan process. In any case it is
a challenge for theorists to study such nonperturbative effects in quantum
field theory. And here is my outlook and the message which I would like to
convey: The {\bf THEORY} of confined quarks may lead to
\begin{center}{\bf M}ore un{\bf E}xpected
ph{\bf E}nomena {\bf T}han {\bf S}tandard \end{center}
\noindent perturbation theory predicts.
{\bf EXPERIMENT}alists
please check.

\bigskip
\noindent{\bf Acknowledgements:} The author would like to thank
all colleagues with whom he had the pleasure to collaborate on the
topics mentioned in this seminar. For many fruitful discussions
he is grateful to H. G. Dosch and P. V. Landshoff.
Special thanks are due to G. W. Botz for his help in the
preparation of the figures for this article. Finally the author would
like to thank R. Casalbuoni, L. Lusanna, and their colleagues at
Florence for providing such a pleasant atmosphere at the workshop.

\newpage
\section*{Figure Captions}
\begin{description}
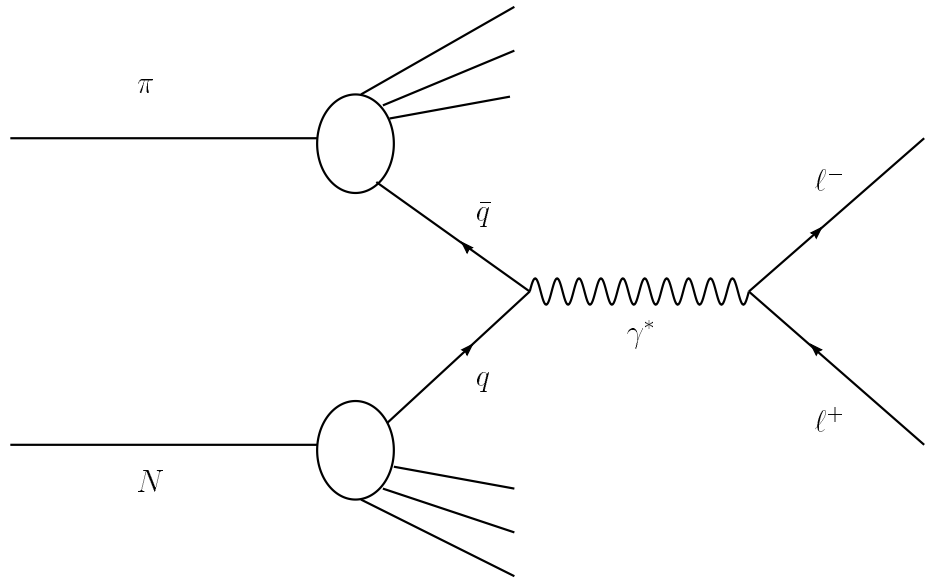
\item[Fig. 1:]   The lowest order diagram for the Drell-Yan reaction
(\ref{1.2})
in the QCD improved parton model.
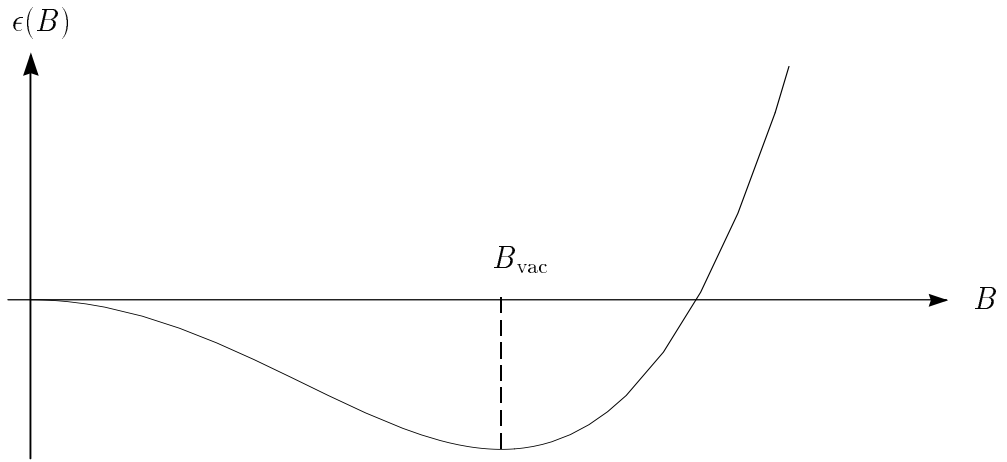
\item[Fig. 2:]  The schematic behaviour of the vacuum energy density
$\varepsilon (B)$ as function of a constant chromomagnetic field $B$ according
to
Savvidy's calculation (eq. (\ref{2.2})).
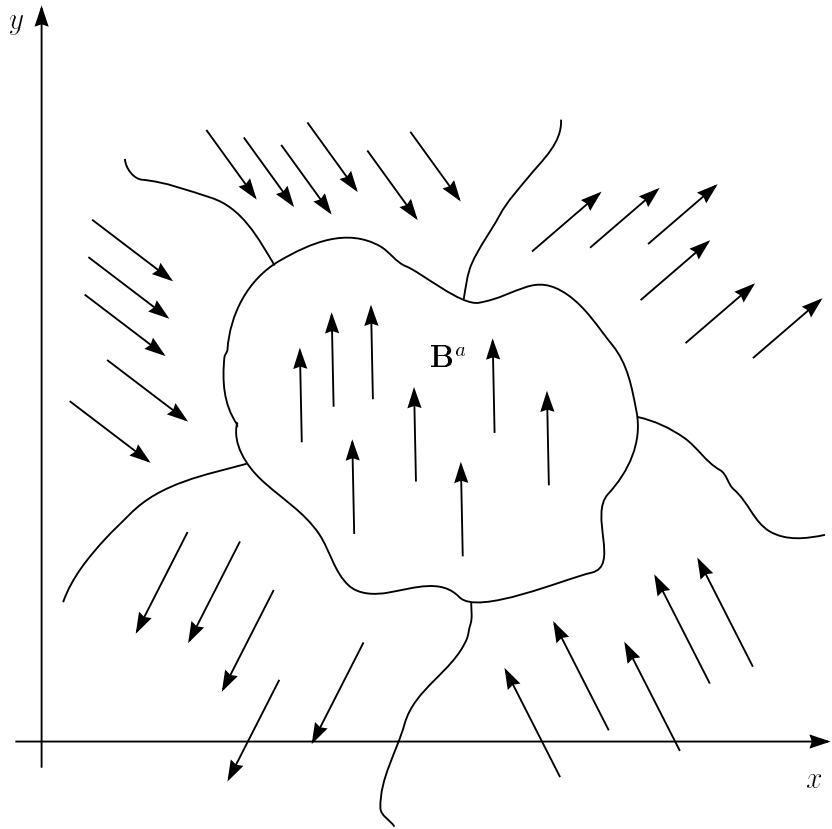
\item[Fig. 3:]  A ``snapshot'' of the QCD vacuum showing a domain structure of
spontaneously created chromomagnetic fields.
\item[Fig. 4:]  The QCD-vacuum according to Ambj\o rn and Oleson \cite{28} (a).
The ether according to Maxwell \cite{33} (b).
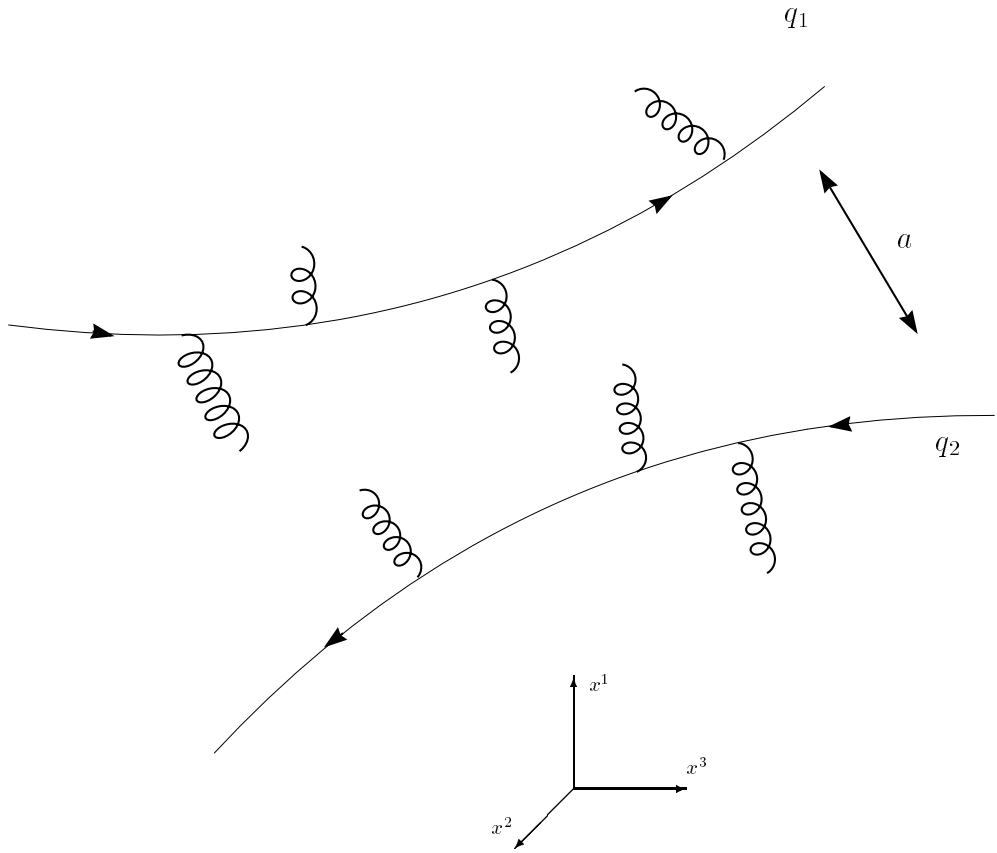
\item[Fig. 5:]  An encounter of two quarks: $q_1$ moving fast in positive and
$q_2$ fast in negative $x^3$ direction.  The interaction of the quarks with the
gluons in the QCD vacuum is indicated by the spiral lines.
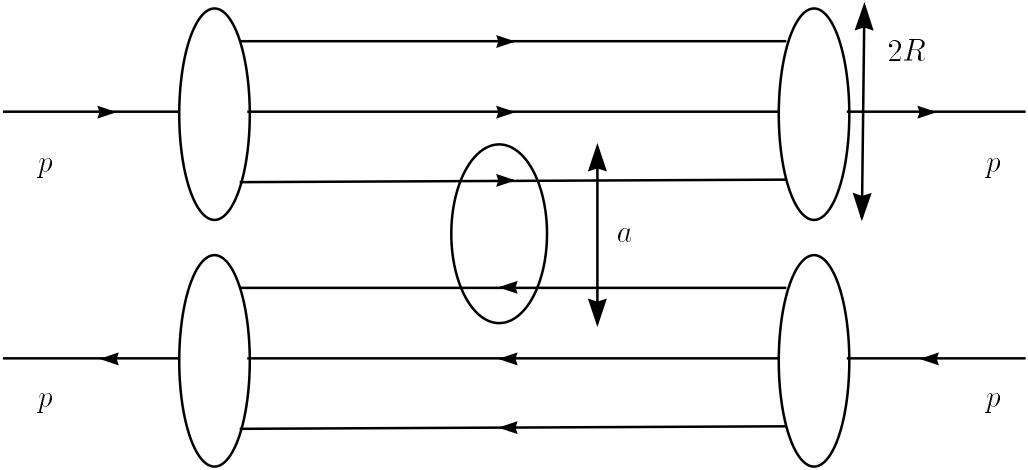
\item[Fig. 6:]  Proton-proton elastic scattering at high energies with the
scattering of single quarks on each other.
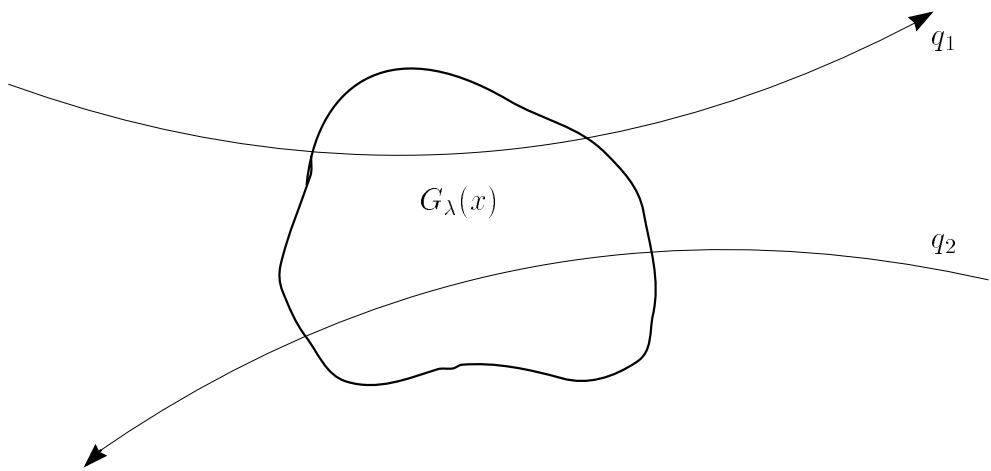
\item[Fig. 7:]  Scattering of two quarks $q_1,q_2$ in a given external gluon
potential $G_\lambda(x)$.
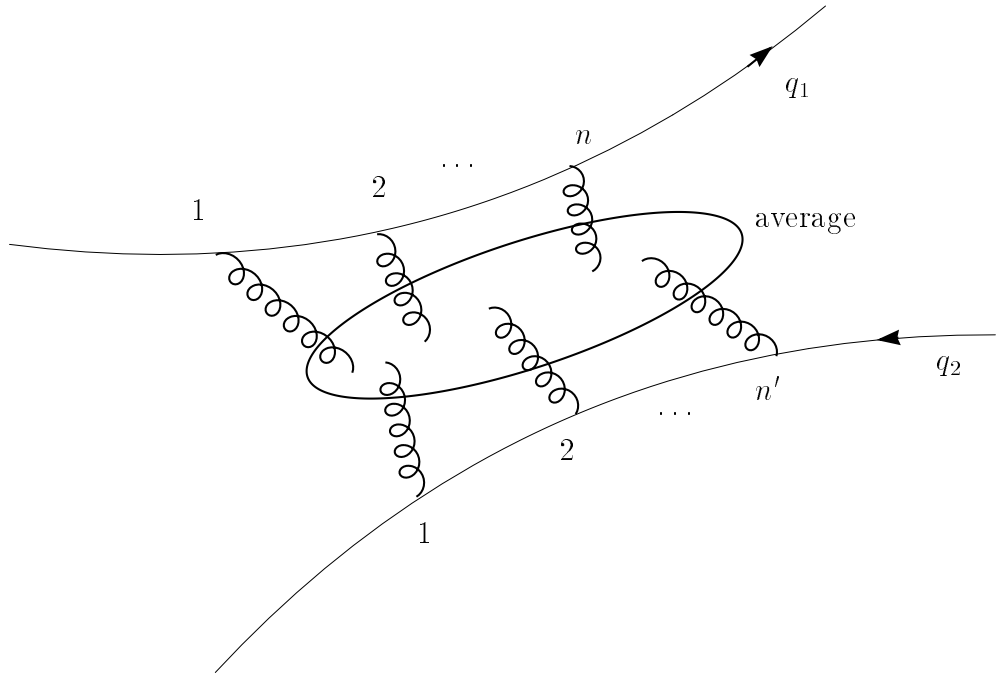
\item[Fig. 8:]  Schematic representation of the functional integral on the
r.h.s. of (\ref{3.4}) which sums up the diagrams with arbitrary numbers
$n$ and $n'$ of gluon vertices on the quark lines $q_1$ and $q_2$,
respectively. This gives the amplitude
 for $q-q$ scattering.
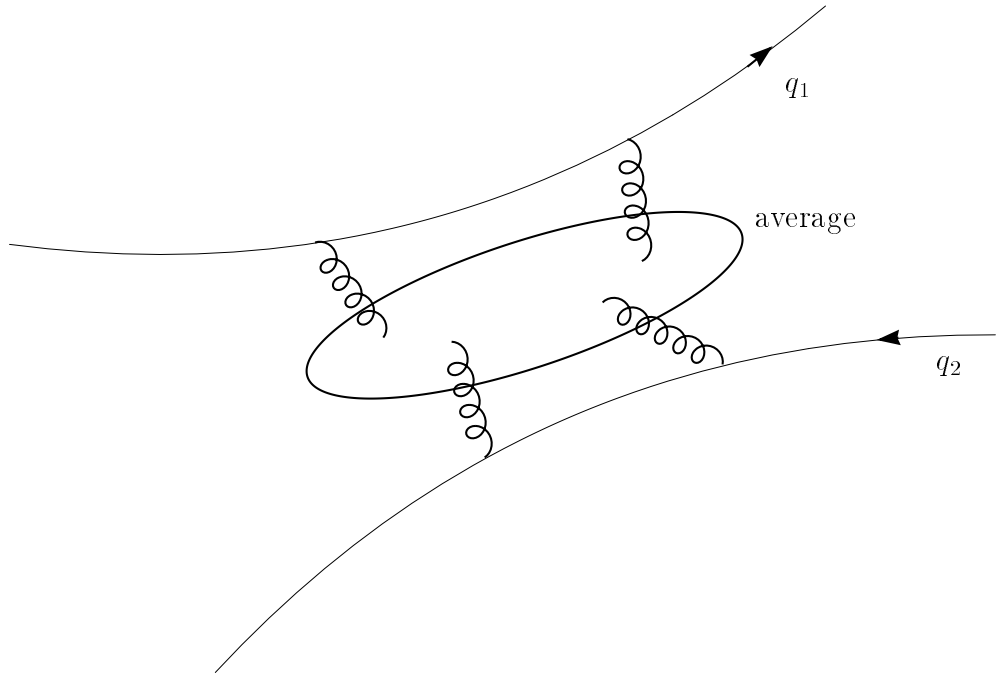
\item[Fig. 9:] Simplest approximation for the functional integral on the
r.h.s. of (\ref{3.4}) for $q-q$ scattering without colour exchange.
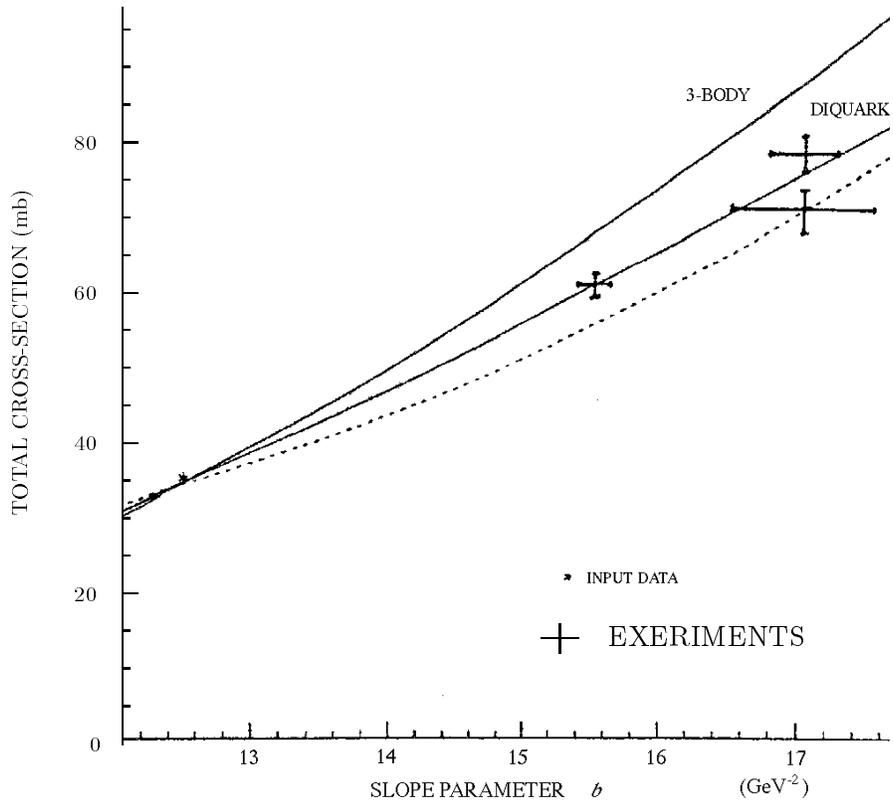
\item[Fig. 10:]  The relation between the total cross section $\sigma_{\rm
tot}$ and the slope parameter $b$ for proton-proton and proton-antiproton
scattering. The dotted line is the prediction  from Regge theory. The
prediction of the calculation for soft high energy scattering in the stochastic
vacuum model is that the data points should lie in the area between the full
lines.
In essence this is given by (\ref{3.8}), (\ref{3.9}) with an
uncertainty estimate from different assumptions for the proton wave functions
(cf.
\cite{41}).
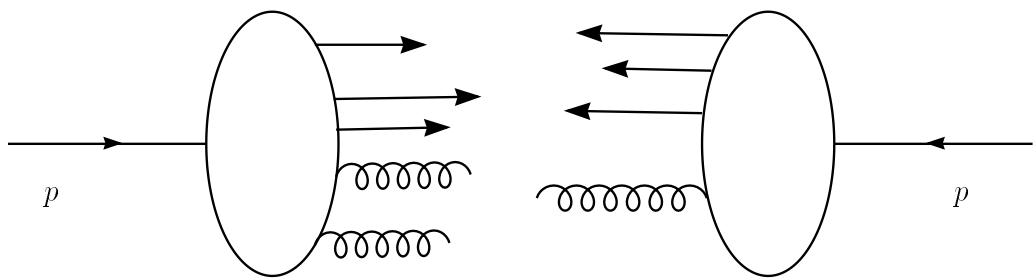
\item[Fig. 11:]   A proton-proton collision at high energies in the
parton picture.
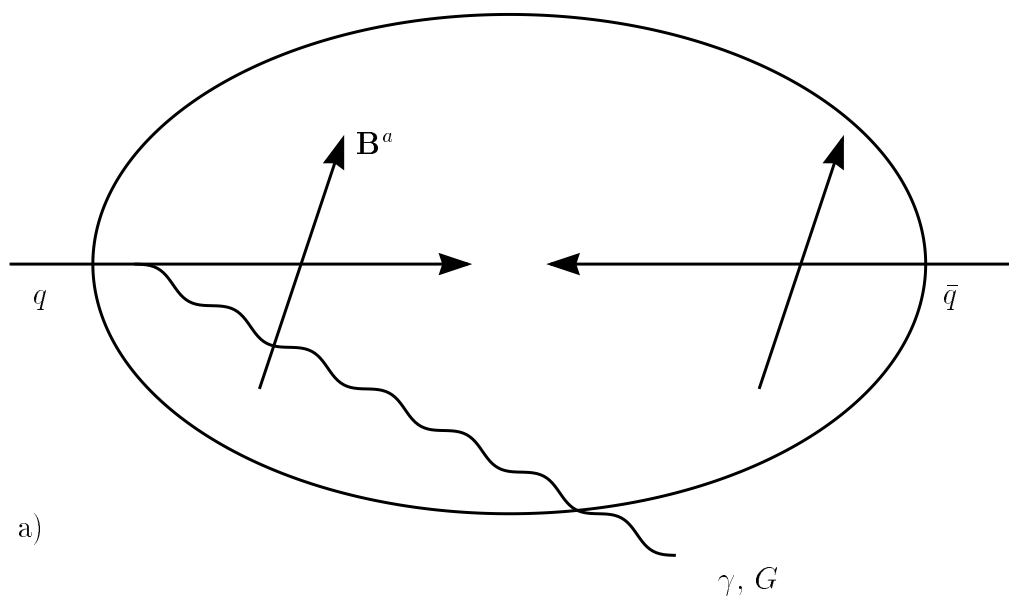
\item[Fig. 12:]   A quark and antiquark traversing a region of chromomagnetic
field (a). An electron and a positron in a storage  ring (b). In both cases we
expect the emission of synchrotron radiation to occur.
\end{description}

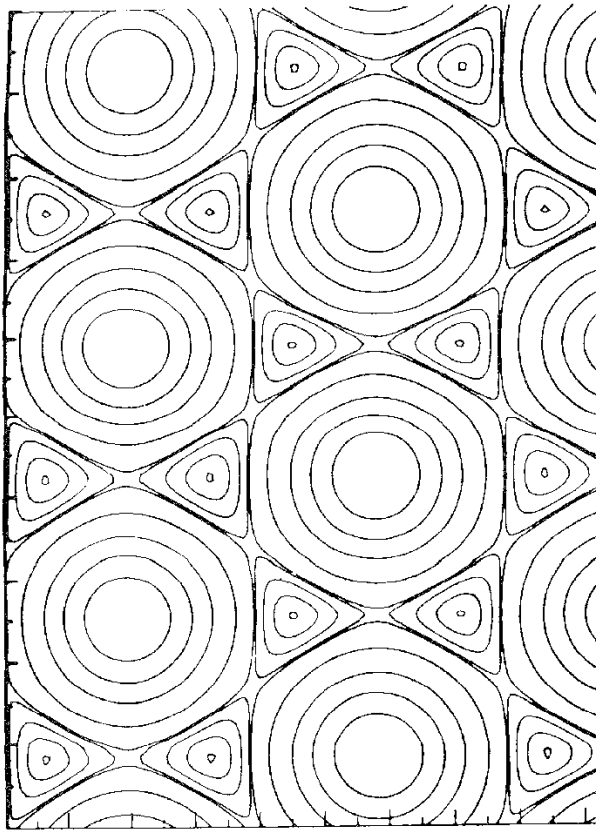
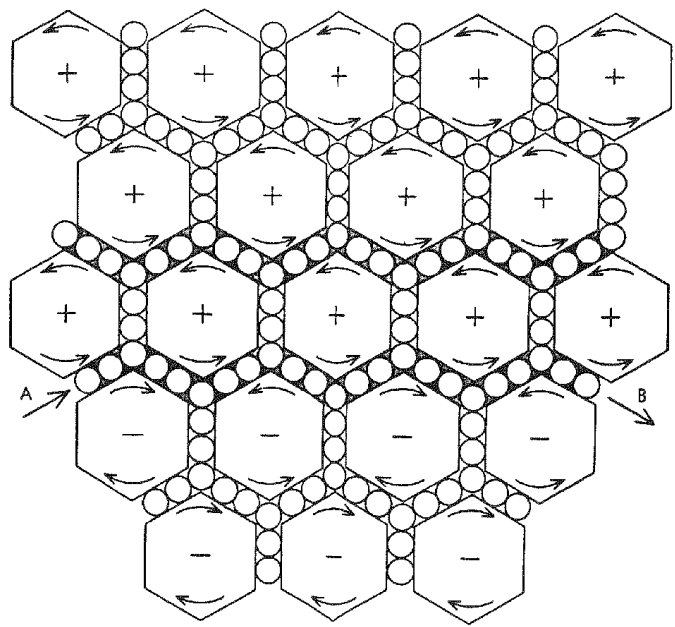
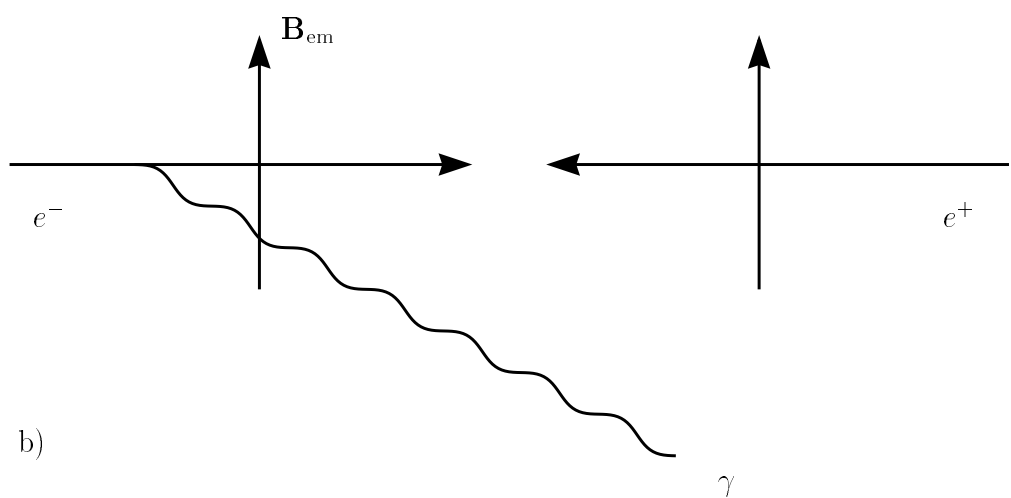
\end{document}